\documentclass[10pt,a4paper,onecolumn]{article}
\usepackage{marginnote}
\usepackage{graphicx}
\usepackage[rgb]{xcolor}
\usepackage{authblk,etoolbox}
\usepackage{titlesec}
\usepackage{calc}
\usepackage{tikz}
\usepackage[pdfa]{hyperref}
\usepackage{hyperxmp}
\hypersetup{%
	unicode=true,
	pdfapart=3,
	pdfaconformance=B,
	pdftitle={RelativisticDynamics.jl: Relativistic Spin-Orbital
		Dynamics in Julia},
	pdfauthor={Tom Kimpson},
	pdfpublication={Journal of Open Source Software},
	pdfpublisher={Open Journals},
	pdfissn={2475-9066},
	pdfpubtype={journal},
	pdfvolumenum={},
	pdfissuenum={},
	pdfdoi={10.xxxxxx/draft},
	pdfcopyright={Copyright (c) 1970, Tom Kimpson},
	pdflicenseurl={http://creativecommons.org/licenses/by/4.0/},
	colorlinks=true,
	linkcolor=[rgb]{0.0, 0.5, 1.0},
	citecolor=Blue,
	urlcolor=[rgb]{0.0, 0.5, 1.0},
	breaklinks=true
}
\usepackage{caption}
\usepackage{orcidlink}
\usepackage{amssymb,amsmath}
\usepackage{ifxetex,ifluatex}
\usepackage{seqsplit}
\usepackage{xstring}

\usepackage{float}
\let\origfigure\figure
\let\endorigfigure\endfigure
\renewenvironment{figure}[1][2] {
	\expandafter\origfigure\expandafter[H]
} {
	\endorigfigure
}

\usepackage{fixltx2e} % provides \textsubscript

\newlength{\cslhangindent}
\newlength{\csllabelwidth}
\newenvironment{CSLReferences}[2] % #1 hanging-ident, #2 entry spacing
{% don't indent paragraphs
	\setlength{\parindent}{0pt}
	\ifodd #1 \everypar{\setlength{\hangindent}{\cslhangindent}}\ignorespaces\fi
	\ifnum #2 > 0
	\setlength{\parskip}{#2\baselineskip}
	\fi
}%
{}
\usepackage{calc}

\usepackage[top=3.5cm, bottom=3cm, right=1.5cm, left=1.0cm,
headheight=2.2cm, reversemp, includemp, marginparwidth=4.5cm]{geometry}

\titleformat{\section}
{\normalfont\sffamily\Large\bfseries}
{}{0pt}{}
\titleformat{\subsection}
{\normalfont\sffamily\large\bfseries}
{}{0pt}{}
\titleformat{\subsubsection}
{\normalfont\sffamily\bfseries}
{}{0pt}{}
\titleformat*{\paragraph}
{\sffamily\normalsize}

\usepackage{fancyhdr}
\makeatletter
\let\ps@plain\ps@fancy
\fancyheadoffset[L]{4.5cm}
\fancyfootoffset[L]{4.5cm}

\definecolor{linky}{rgb}{0.0, 0.5, 1.0}

\newcommand{\ExternalLink}{%
	\tikz[x=1.2ex, y=1.2ex, baseline=-0.05ex]{%
		\begin{scope}[x=1ex, y=1ex]
			\clip (-0.1,-0.1)
			--++ (-0, 1.2)
			--++ (0.6, 0)
			--++ (0, -0.6)
			--++ (0.6, 0)
			--++ (0, -1);
			\path[draw,
			line width = 0.5,
			rounded corners=0.5]
			(0,0) rectangle (1,1);
		\end{scope}
		\path[draw, line width = 0.5] (0.5, 0.5)
		-- (1, 1);
		\path[draw, line width = 0.5] (0.6, 1)
		-- (1, 1) -- (1, 0.6);
	}
}

\patchcmd{\@maketitle}{center}{flushleft}{}{}
\patchcmd{\@maketitle}{center}{flushleft}{}{}
\patchcmd{\@maketitle}{\LARGE}{\LARGE\sffamily}{}{}
\def\maketitle{{%
		
		\AB@maketitle}}
\renewcommand\AB@affilsepx{ \protect\Affilfont}
\renewcommand\AB@affilnote[1]{{\bfseries #1}\hspace{3pt}}
\renewcommand{\affil}[2][]%
{\newaffiltrue\let\AB@blk@and\AB@pand
	\if\relax#1\relax\def\AB@note{\AB@thenote}\else\def\AB@note{#1}%
	\setcounter{Maxaffil}{0}\fi
	\begingroup
	\let\href=\href@Orig
	\let\protect\@unexpandable@protect
	\def\thanks{\protect\thanks}\def\footnote{\protect\footnote}%
	\@temptokena=\expandafter{\AB@authors}%
	{\def\\{\protect\\\protect\Affilfont}\xdef\AB@temp{#2}}%
	\xdef\AB@authors{\the\@temptokena\AB@las\AB@au@str
		\protect\\[\affilsep]\protect\Affilfont\AB@temp}%
	\gdef\AB@las{}\gdef\AB@au@str{}%
	{\def\\{, \ignorespaces}\xdef\AB@temp{#2}}%
	\@temptokena=\expandafter{\AB@affillist}%
	\xdef\AB@affillist{\the\@temptokena \AB@affilsep
		\AB@affilnote{\AB@note}\protect\Affilfont\AB@temp}%
	\endgroup
	\let\AB@affilsep\AB@affilsepx
}
\makeatother

\renewcommand\Affilfont{\sffamily\small\mdseries}
\IfFileExists{upquote.sty}{\usepackage{upquote}}{}
\IfFileExists{microtype.sty}{%
	\usepackage{microtype}
	\UseMicrotypeSet[protrusion]{basicmath} % disable protrusion for tt fonts
}{}

\PassOptionsToPackage{usenames,dvipsnames}{color} % color is loaded by hyperref
\ifLuaTeX
\usepackage[bidi=basic]{babel}
\else
\usepackage[bidi=default]{babel}
\fi
\babelprovide[main,import]{american}
\def\languageshorthands#1{}
\usepackage{lineno}
\usepackage{graphicx,grffile}
\makeatletter
\def\maxwidth{\ifdim\Gin@nat@width>\linewidth\linewidth\else\Gin@nat@width\fi}
\def\maxheight{\ifdim\Gin@nat@height>\textheight\textheight\else\Gin@nat@height\fi}
\makeatother
\setkeys{Gin}{width=\maxwidth,height=\maxheight,keepaspectratio}
\IfFileExists{parskip.sty}{%
	\usepackage{parskip}
}{% else
	\setlength{\parindent}{0pt}
	\setlength{\parskip}{6pt plus 2pt minus 1pt}
}
\ifx\paragraph\undefined\else
\let\oldparagraph\paragraph
\renewcommand{\paragraph}[1]{\oldparagraph{#1}\mbox{}}
\fi
\ifx\subparagraph\undefined\else
\let\oldsubparagraph\subparagraph
\renewcommand{\subparagraph}[1]{\oldsubparagraph{#1}\mbox{}}
\fi
\ifLuaTeX
\usepackage{selnolig}  % disable illegal ligatures
\fi

\title{RelativisticDynamics.jl: Relativistic Spin-Orbital Dynamics in
	Julia}

\author[1,2%
]{Tom Kimpson%
}

\affil[1]{School of Physics, University of Melbourne, Parkville, VIC
	3010, Australia}
\affil[2]{Australian Research Council (ARC) Centre of Excellence for
	Gravitational Wave Discovery (OzGrav)}
\date{\vspace{-2.5ex}}

\begin{document}
	\maketitle
	
	\marginpar{
		
		\begin{flushleft}
			%\hrule
			\sffamily\small
			
			{\bfseries DOI:} \href{https://doi.org/10.21105/joss.04992}{\color{linky}{10.21105/joss.04992}}
			
			\vspace{2mm}
			{\bfseries Software}
			\begin{itemize}
				\setlength\itemsep{0em}
				\item \href{https://github.com/openjournals/joss-reviews/issues/4992}{\color{linky}{Review}} \ExternalLink
				\item \href{https://github.com/tomkimpson/RelativisticDynamics.jl}{\color{linky}{Repository}} \ExternalLink
				\item \href{https://doi.org/10.5281/zenodo.8412240}{\color{linky}{Archive}} \ExternalLink
			\end{itemize}
			
			\vspace{2mm}
			
			\par\noindent\hrulefill\par
			
			\vspace{2mm}
			
%			{\bfseries Editor:} \href{https://joss.theoj.org}{Open
%				Journals} \ExternalLink
%			\\
%			\vspace{1mm}
%			{\bfseries Reviewers:}
%			\begin{itemize}
%				\setlength\itemsep{0em}
%				\item \href{https://github.com/openjournals}{@openjournals}
%			\end{itemize}
			\vspace{2mm}
			
			{\bfseries Submitted:} 15 November 2022\\
			{\bfseries Published:} 10 October 2023
			
			\vspace{2mm}
			{\bfseries License}\\
			Authors of papers retain copyright and release the work under a Creative Commons Attribution 4.0 International License (\href{https://creativecommons.org/licenses/by/4.0/}{\color{linky}{CC BY 4.0}}).

		\end{flushleft}
	}
	
	\hypertarget{summary}{%
		\section{Summary}\label{summary}}
	
	Relativistic binaries composed of a millisecond pulsar (MSP) orbiting a
	much more massive (\(\gtrsim 10^3 M_{\odot}\)), spinning black hole (BH)
	are exceptional probes for investigating key questions of fundamental
	physics and astrophysics. Such systems are natural sources of
	gravitational waves (GWs) in the mHz regime, expected to be detectable
	by the next generation of space-based GW detectors such as LISA
	(\protect\hyperlink{ref-LISA}{Thorpe et al., 2019}). The associated
	radio emission from the companion pulsar raises the possibility of an
	electromagnetic (EM) counterpart, enabling high precision multimessenger
	measurements to be made. The description of the orbital dynamics of
	these systems, and the influence on the resultant observed EM and GW
	signals, is non-trivial. A proper treatment of the spin-orbital dynamics
	can be derived from the conservation of the energy-momentum tensor
	\begin{equation}\label{eq:conservation}
		{T^{\mu \nu}}_{;\nu} = 0
	\end{equation} which when expanded into a set of infinite multipole
	moments leads to a description of the momentum vector \(p^{\mu}\) and
	the spin tensor \(s^{\mu \nu}\) \begin{equation}\label{eq:mpd1}
		\frac{Dp^{\mu}}{d \lambda} = -\frac{1}{2}{R^{\mu}}_{\nu \alpha \beta} u^{\nu} s^{\alpha \beta}
	\end{equation} \begin{equation}\label{eq:mpd2}
		\frac{Ds^{\mu \nu}}{d \lambda} =p^{\mu}u^{\nu} - p^{\nu}u^{\mu}
	\end{equation} for affine parameter \(\lambda\), 4-velocity \(u^{\nu}\)
	and Riemann curvature tensor \({R^{\mu}}_{\nu \alpha \beta}\). The
	system is closed by providing a spin supplementary condition, equivalent
	to specifying the observer-dependent centre of mass. For this work we
	take the Tulczyjew-Dixon condition
	(\protect\hyperlink{ref-Dixon1964}{Dixon, 1964};
	\protect\hyperlink{ref-tulczyjew}{Tulczyjew, 1959})
	\begin{equation}\label{eq:mpd3}
		s^{\mu \nu} p_{\nu} = 0
	\end{equation}
	
	Together, equations \ref{eq:mpd1} - \ref{eq:mpd3} form the
	Mathisson-Papetrou-Dixon (MPD) equations
	(\protect\hyperlink{ref-Dixon1964}{Dixon, 1964};
	\protect\hyperlink{ref-Mathisson1937}{Mathisson, 1937};
	\protect\hyperlink{ref-Papapetrou1951}{Papapetrou, 1951}), and describe
	the spin-orbital evolution in a fully consistent way that is applicable
	to strong field regimes.
	
	\hypertarget{statement-of-need}{%
		\section{Statement of need}\label{statement-of-need}}
	
	\texttt{RelativisticDynamics.jl} is an open-source Julia package for
	relativistic spin-orbital dynamics in the gravitational strong field for
	a Kerr spacetime. Existing codes for modelling the dynamics of spinning
	objects like pulsars in the strong-field regime are generally lacking,
	since such systems occupy an intermediate regime that is generally
	overlooked. At the “low” end of this regime there are post-Newtonian or
	geodesic descriptions (e.g.
	\protect\hyperlink{ref-PhysRevD.45.1840}{Damour \& Taylor, 1992}) which
	neglect the influence of the pulsar spin on the underlying spacetime
	metric (“spin-curvature” coupling). At the “high” end there is the full
	Numerical Relativity (NR) solutions (e.g.
	\protect\hyperlink{ref-Andrade2021}{Andrade et al., 2021}) which are
	primarily applicable to two BHs with a mass ratio \(\mathcal{O}(1)\),
	and are computationally intractable for these MSP systems which are
	observed over a large number of orbital cycles.
	
	\texttt{RelativisticDynamics.jl} aims to bridge this gap by providing a
	modern, fast code for accurate numerical evolution of spinning
	relativistic systems, via the MPD formalism. Julia is a modern language
	that solves the “two language problem”, enabling fast dynamic typing and
	JIT compilation in conjunction with petaflop performance, comparable
	with numerical languages that are better known in the astrophysics
	community such as C or Fortran. As a modern language, it also provides a
	dedicated package manager and a large catalogue of \emph{composable}
	packages for scientific computing. This enables
	\texttt{RelativisticDynamics.jl} to easily leverage and interface with
	other scientific computing packages. The author and collaborators have
	used the general methods and mathematics described in this package for
	multiple research projects (e.g.
	\protect\hyperlink{ref-Kimpson2019}{Kimpson et al., 2019},
	\protect\hyperlink{ref-Kimpson2020}{2020a},
	\protect\hyperlink{ref-KimpsonAA}{2020b};
	\protect\hyperlink{ref-Li2019}{Li et al., 2019}) with a particular focus
	on the radio signals from spinning pulsar systems. This package
	represents an attempt to create a documented, well-tested, open source
	resource for public use in this area, that can also be used as a
	computational playground for exploring techniques that could be
	applicable to more advanced numerical models. The package has been
	formulated in terms of ODE integration, rather than using
	e.g.~action-angle variables
	(\protect\hyperlink{ref-Witzany2022}{Witzany, 2022}), to allow for
	extension to general spacetime metrics and straightforward computation
	of quantities relevant for pulsar observations e.g.~spin axis
	orientation.
	
	In addition to providing a fast, modern package for strong field spin
	dynamics, \texttt{RelativisticDynamics.jl} has two additional important
	features from the perspective of modern relativistic astrophysics.
	Firstly, it is fully type flexible, being able to support arbitrary
	number formats. By making use of Julia’s type-flexibility the model is
	written in such a way so as to be able to support hardware accelerated,
	low precision arithmetic and alternative rounding methods such as
	stochastic rounding. This enables rapid prototyping and exploration of
	reduced precision numerical techniques in astrophysics, an approach
	common in other numerical fields such as weather and climate modelling
	(e.g. \protect\hyperlink{ref-ECMWF}{Váňa et al., 2017}). Secondly,
	\texttt{RelativisticDynamcis.jl} is written to be fully differentiable
	via automatic differentiation. This enables the package to be used for
	differentiable physics applications in astrophysics, for example
	gravitational waveform modelling and parameter estimation or training
	neural networks based on the model. Automatic differentiation also
	provides a potential avenue for extension of the package to general
	(i.e.~non-Kerr) spacetimes, whereby a user can specify the metric and
	the associated Christoffel symbols and Riemann tensors - which are
	simply linear combinations of the metric derivatives - are calculated
	automatically.
	
	Future potential extensions of this code include taking the dynamics
	beyond second order in the multipole expansion, and the inclusion of
	alternative spin conditions and spacetime metrics. The inclusion of a
	diagnostics tool for extracting gravitational waveforms in the time
	domain via a numerical kludge method would also be a worthwhile
	addition. Moreover, we have considered only bound dynamical systems -
	the ability to also explore hyberbolic systems would also be an
	interesting development.
	
	\includegraphics[width=0.5\textwidth,height=\textheight]{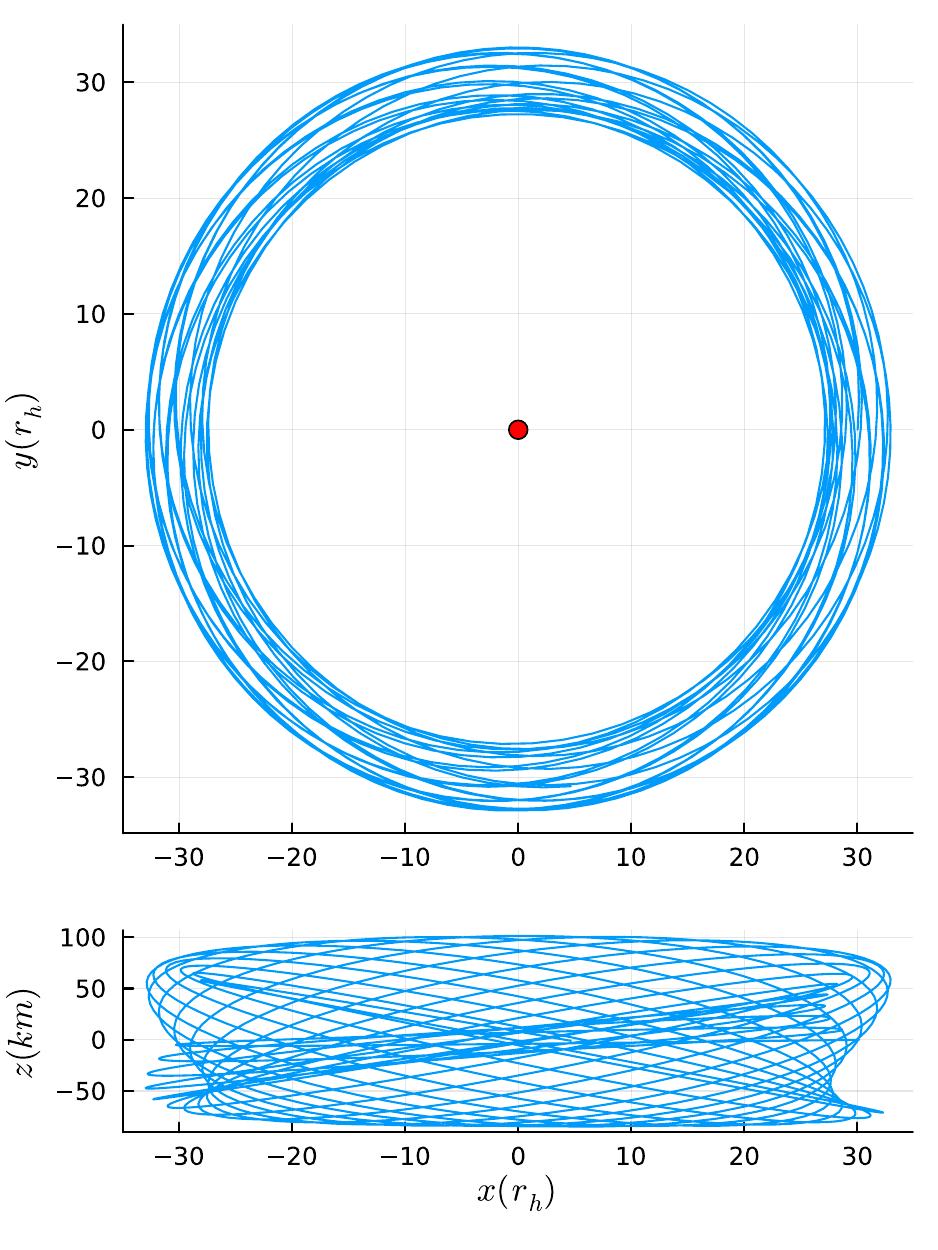}
	\includegraphics[width=0.5\textwidth,height=\textheight]{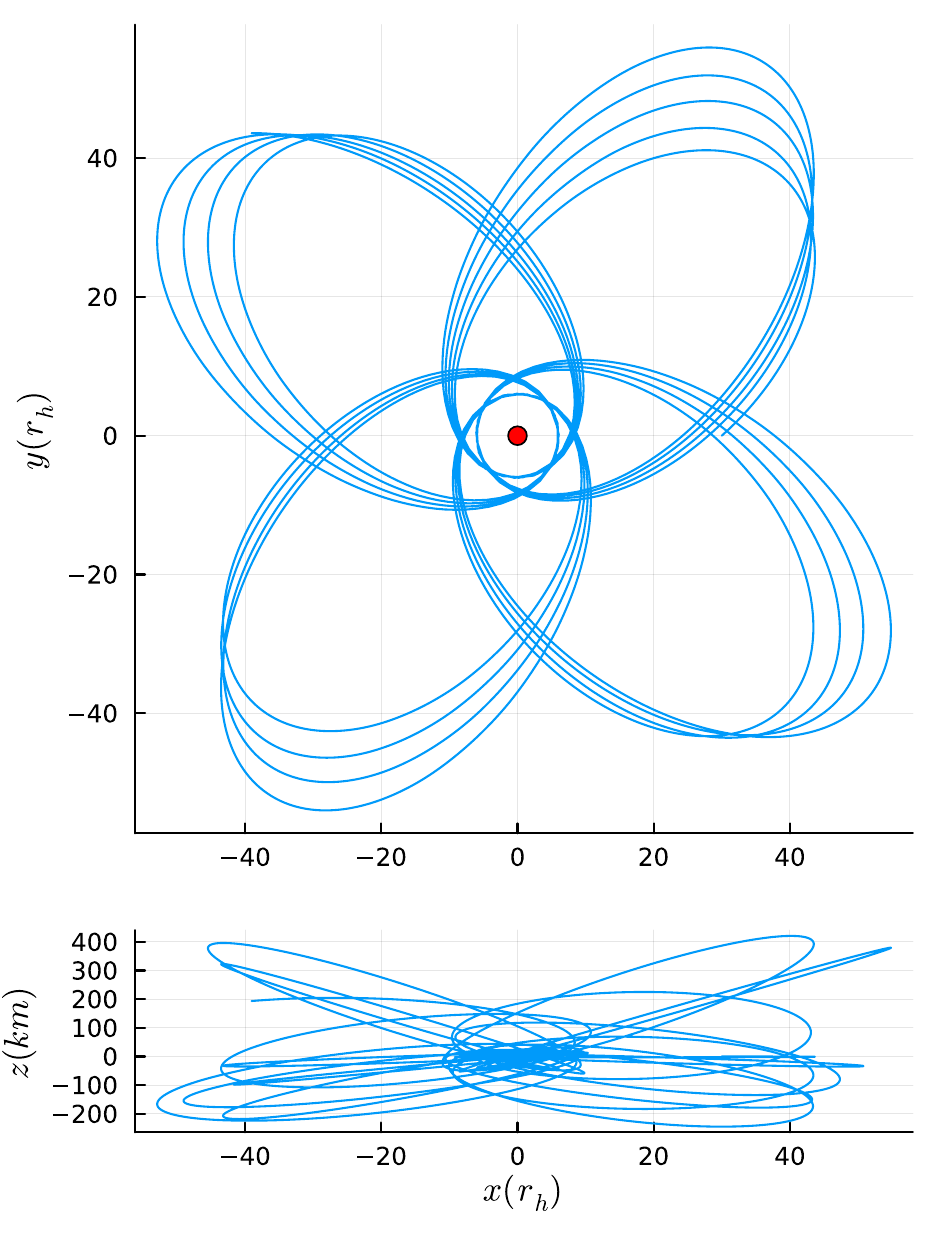}
	
	\begin{figure}[!h]
		\caption{Example orbital trajectories for a ms-pulsar with eccentricity $e=0.1$ (left panels), $e=0.8$ (right panels), orbiting a massive BH with extremal spin, $a=0.998$. The orbital motion is presented in the $x-y$ plane (top panels) and $x-z$ plane (bottom panels). The pulsar is initialised in the orbital plane with zero inclination. In the absence of spin-curvature coupling the particle would remain in the plane ($z=0$). Note the $z$-motion is on the scale of km, not gravitational radii.}
	\end{figure}
	
	\hypertarget{acknowledgements}{%
		\section{Acknowledgements}\label{acknowledgements}}
	
	This work exploring the spin-evolution of relativistic systems via the
	the MPD equations was originally motivated through interesting
	discussions with Kinwah Wu. The port to a modern, precision-flexible
	model in Julia was heavily inspired by Milan Klöwer. Our thanks to both.
	
	\hypertarget{references}{%
		\section*{References}\label{references}}
	\addcontentsline{toc}{section}{References}
	
	\hypertarget{refs}{}
	\begin{CSLReferences}{1}{0}
		\leavevmode\vadjust pre{\hypertarget{ref-Andrade2021}{}}%
		Andrade, T., Salo, L. A., Aurrekoetxea, J. C., Bamber, J., Clough, K.,
		Croft, R., Jong, E. de, Drew, A., Duran, A., Ferreira, P. G., Figueras,
		P., Finkel, H., França, T., Ge, B.-X., Gu, C., Helfer, T., Jäykkä, J.,
		Joana, C., Kunesch, M., … Wong, K. (2021). GRChombo: An adaptable
		numerical relativity code for fundamental physics. \emph{Journal of Open
			Source Software}, \emph{6}(68), 3703.
		\url{https://doi.org/10.21105/joss.03703}
		
		\leavevmode\vadjust pre{\hypertarget{ref-PhysRevD.45.1840}{}}%
		Damour, T., \& Taylor, J. H. (1992). Strong-field tests of relativistic
		gravity and binary pulsars. \emph{Phys. Rev. D}, \emph{45}, 1840–1868.
		\url{https://doi.org/10.1103/PhysRevD.45.1840}
		
		\leavevmode\vadjust pre{\hypertarget{ref-Dixon1964}{}}%
		Dixon, W. G. (1964). {A covariant multipole formalism for extended test
			bodies in general relativity}. \emph{Il Nuovo Cimento}, \emph{34}(2),
		317–339. \url{https://doi.org/10.1007/BF02734579}
		
		\leavevmode\vadjust pre{\hypertarget{ref-Kimpson2019}{}}%
		Kimpson, T., Wu, K., \& Zane, S. (2019). {Pulsar timing in extreme mass
			ratio binaries: a general relativistic approach}. \emph{486}(1),
		360–377. \url{https://doi.org/10.1093/mnras/stz845}
		
		\leavevmode\vadjust pre{\hypertarget{ref-Kimpson2020}{}}%
		Kimpson, T., Wu, K., \& Zane, S. (2020a). {Orbital spin dynamics of a
			millisecond pulsar around a massive BH with a general mass quadrupole}.
		\emph{Monthly Notices of the RAS}, \emph{497}(4), 5421–5431.
		\url{https://doi.org/10.1093/mnras/staa2103}
		
		\leavevmode\vadjust pre{\hypertarget{ref-KimpsonAA}{}}%
		Kimpson, T., Wu, K., \& Zane, S. (2020b). {Radio timing in a millisecond
			pulsar - extreme/intermediate mass ratio binary system}. \emph{Astronomy
			and Astrophysics}, \emph{644}, A167.
		\url{https://doi.org/10.1051/0004-6361/202038561}
		
		\leavevmode\vadjust pre{\hypertarget{ref-Li2019}{}}%
		Li, K. J., Wu, K., \& Singh, D. (2019). {Spin dynamics of a millisecond
			pulsar orbiting closely around a massive black hole}. \emph{Monthly
			Notices of the Royal Astronomical Society}, \emph{485}(1), 1053–1066.
		\url{https://doi.org/10.1093/mnras/stz389}
		
		\leavevmode\vadjust pre{\hypertarget{ref-Mathisson1937}{}}%
		Mathisson, A. (1937). \emph{Acta Phys. Pol.}, \emph{6}, 163.
		\url{https://doi.org/10.1007/s10714-010-0939-y}
		
		\leavevmode\vadjust pre{\hypertarget{ref-Papapetrou1951}{}}%
		Papapetrou, A. (1951). Spinning test-particles in general relativity. i.
		\emph{Proceedings of the Royal Society of London A: Mathematical,
			Physical and Engineering Sciences}, \emph{209}(1097), 248–258.
		\url{https://doi.org/10.1098/rspa.1951.0200}
		
		\leavevmode\vadjust pre{\hypertarget{ref-LISA}{}}%
		Thorpe, J. I., Ziemer, J., Thorpe, I., Livas, J., Conklin, J. W.,
		Caldwell, R., Berti, E., McWilliams, S. T., Stebbins, R., Shoemaker, D.,
		Ferrara, E. C., Larson, S. L., Shoemaker, D., Key, J. S., Vallisneri,
		M., Eracleous, M., Schnittman, J., Kamai, B., Camp, J., … Wass, P.
		(2019). {The Laser Interferometer Space Antenna: Unveiling the
			Millihertz Gravitational Wave Sky}. \emph{Bulletin of the American
			Astronomical Society}, \emph{51}, 77.
		\url{https://arxiv.org/abs/1907.06482}
		
		\leavevmode\vadjust pre{\hypertarget{ref-tulczyjew}{}}%
		Tulczyjew, W. M. (1959). Motion of multipole particles in general
		relativity theory binaries. \emph{Acta Phys. Pol.}, \emph{18}, 393.
		
		\leavevmode\vadjust pre{\hypertarget{ref-ECMWF}{}}%
		Váňa, F., Düben, P., Lang, S., Palmer, T., Leutbecher, M., Salmond, D.,
		\& Carver, G. (2017). Single precision in weather forecasting models: An
		evaluation with the IFS. \emph{Monthly Weather Review}, \emph{145}(2),
		495–502. \url{https://doi.org/10.1175/MWR-D-16-0228.1}
		
		\leavevmode\vadjust pre{\hypertarget{ref-Witzany2022}{}}%
		Witzany, V. (2022). {Action-angle coordinates for black-hole geodesics
			I: Spherically symmetric and Schwarzschild}. \emph{arXiv e-Prints},
		arXiv:2203.11952. \url{https://doi.org/10.48550/arXiv.2203.11952}
		
	\end{CSLReferences}
	
\end{document}